# Designing And Analysis Issues For An Attack Resilient and Adaptive Medium Access Control Protocol for Computer Networks: An Exclusive Survey

Piyush Kumar Shukla,
Dept. of CSE, UIT, RGPV,
Bhopal, M.P., INDIA,
pphdw@yahoo.com

S. Silakari, Dept. of CSE, UIT,
RGPV, Bhopal, M.P., INDIA,
ssilakari@yahoo.com

S.S. Bhadouria, Dept. of EC,
MITS, Gwalior,
Saritamts61@yahoo.co.in

**Abstract-** Now a days we are trying to improve through put and efficiency of existing MAC protocols used, suffering from different problem like Request to send and clear to send message propagation, and channel chapter effect or Unfairness due to using Binary Exponential Backoff algorithm, effect of noise in communication path, but it may be Possible to cause an attack by Manipulating traffic in the MAC layer and propagate attack to the routing layer then we are in a dilemma how to improve performance of MAC layer protocols . Differentiation of attack and congestion in wireless network is a big issue. We promote and prefer a dynamic contention window size for utilizing unused slots but dynamic behavior increases uncertainty and randomness in protocol behavior and also uneasiness to detecting attacks. It may be possible that a node is maliciously increasing the data traffic into the network, and also affecting the backoff process, it is mandatory to learn pattern the pattern of the attack so that we can discard him as early as possible. Behavior of Malicious node during attack is also burning issue and how much performance of MAC protocol degraded as per increasing the number of Malicious nodes. At what percentage of total nodes involve in any communication.

We have to identify the parameter on which we can work to strengthen MAC layer protocol against attacks.

*Keywords- Attack, Wireless Sensor Network, Z-MAC, S-MAC, MAC, Collision, BEB*

## 1. Introduction

If we have to improve Performance of MAC layer protocol than we have to work in such a way that routing information is always available to MAC layer and we are able to sense falls packets, we have to modified existing protocol or develop a new protocol that can consider this problem and can discard false packets or can modify contention window after immediate information of these packets/detection of malicious node.

Its the time of failure of exercise have done to decide minimum and maximum contention window size for different traffic conditions when attacker utilize or misuse the unused available slots . although duration of the effect of attacks is depend on traffic and obviously on the ratio of malicious packets and on actual packets. It may be possible that we are not affected by attacks when its ratio is near to 0.0

If X is the actal packets and Y is malicious packets that we can derive different conditions to evaluate performance of MAC protocol (Z) during attacks.

Z=Y/X.. Can be evaluated and results can be a deciding factor about when and how to increase Contention window size.

How to achieve better performance with Attack Resilience MAC Protocol/ procedure to built up an atmosphere for Resilience protocol

Due to innovations in the technology and due to rapid growth in the field of Computer Network and Internet, The role of Computer Network [1] for users to access the Internet has become more important. If we have a broadcast network [2][ to access the channel, the key issue is how to determine who gets to use the channel, when there is a competition for it. Broadcast channels are sometimes referred to as Multi Access Channels or Random Access Channels. It is based on a particular Medium Access Control (MAC) [3, 6] protocol used extensively in providing network access for a collection of end users. These protocols belong to a sub-layer of the data link layer called the MAC sub-layer.

Any varying channel allocation protocol or scheme can be apply, tries to get better utilization when the traffic is unpredictable or fluctuating. In contention resolution approaches, users transmit a packet when they have data to send, if multiple users transmit at the same time a collision occurs and the packets must be retransmitted according to some rule. Random access protocols are different from other types of MAC protocols in one property; a transmission is not guaranteed to be a successful one.

In random access or contention methods, no station is superior to another station and none is assigned the control over another. At each instance, a station that has data to send uses a procedure defined by the protocol to make a decision on whether or not to send. This decision depends on the state of the medium (idle or busy). In other words, each station can transmit when it desires on the condition [4] that it follows the predefined procedure, including the testing of the state of the medium.

Two features give this method its name. First, there is no scheduled time for a station to transmit. Transmission is random among the stations. That is why these methods are called random access. Second, no rules specify which station should sent next. Stations compete with one another to access the medium. That is why these methods are also called contention methods.

In the case of a collision, retransmissions are required for all unsuccessful transmissions until they are successful this collision may be increases due to attack. To avoid further collisions of the same set of collided transmissions, the retransmission of an unsuccessful transmission is scheduled in a randomly chosen future time hoping that other retransmissions will not be scheduled to the same future time and simultaneously we have to learn the pattern of packets due to malicious attacks [22, 24, 33]. This leads to the development of retransmission algorithms for random access protocols. Random Access Systems are relatively insensitive to the number of active stations in the system so their performance is best when there are a large number of lightly loaded sources. However their performance degrade in the case of malicious attacks which include Denial of Service Attack, Replay Attack, Continuous Channel Access or Exhaustion attack, Flooding attack, Jamming (Radio interference) attack, Selective forwarding attack. The need of a MAC protocol with the ability to identify threats and capability to discard these useless packet is required otherwise performance of shared medium may be degraded. We Focus Our Efforts on further improving the channel allocation scheme and include increase in data traffic due to various attacks, we can adopt a Pattern matching approach to learn and mitigate the attacks [28]. An Efficient and Adaptive Medium Access Control Scheme using Pattern matching from the Attacks can be design to improve the performance of Media Access Control protocol.

## 2. Analysis of work done to explain attacks on networks

In this author said that There are basically three levels of DoS attacks [1], growing both in sophistication and seriousness of attack effects. The simplest attack exploits errors and bugs in the design and source code of a network operating system. The second level of attack exploits known artifacts of a particular system implementation or protocol, often due to limited storage or capacity, to introduce delay, to saturate a system, or otherwise limit accessibility. The third and most damaging level of attack uses very specific features of the network protocol to mount the attack. These attacks are specifically designed to look like normal usage. Protecting against DoS attacks is a difficult and complex problem. There is no single approach or answer to increasing our resiliency and protection against them. In fact, the solution will likely be a combination of approaches, including optimizing server settings, using network topologies that minimize attack effects, and sweeping changes to protocols and router behaviour.

In this it is explained that Attack model is the foundation for organizing and implementing attacks [2] against the target system in Attack Resistance Test. By redefining the node of the attack tree model and re-describing the relation of the attack tree nodes, we build a penetration attack tree model which can describe, organize, classify, manage and schedule the attacks for Attack Resistance Test. The organization method of the penetration attack tree is designed in this paper, and an algorithm of attack serialization is put forward. We also design and realize a penetration attack system whose attack scheme is the instance of the model. In the end we present an execution example of the penetration attack system. The example shows that the penetration attack model can describe the logical relationship of the attacks detailed and effectively, and its serialization result can provide the guidance for penetration attack. Authors said that in the multimedia applications over the mobile ad hoc networks, the goodput and delay performance of UDP packets is very sensitive to the congestion targeted DDoS attacking. this paper, they analyze this type of attacking in details for the first time to our knowledge. they figure out the principles of the attacking from the analysis of network capacity and classify the attacking into four categories: Pulsing attacking, Round Robin attacking, Self-whisper attacking and Flooding attacking.

The defense schemes are also proposed, which includes the detection of RTS/CTS packets, signal interference frequency and retransmission time and response stage with ECN marking mechanism. Survey of the security in mobile ad hoc networks are given in paper [6], however, they discuss more about secure routing protocols and key management schemes, and few concerns of the MAC DDoS attacking behaviors. Y. Zhang et al [7] figure out there are different possible attacks to the different layer in the protocol stack, however, they leave the description of detail attacking patterns and DDoS attacking in the future works. V. Gupta et al [8] points out there are attacking to route layer and MAC layer even within the end-to-end authentication network. If there are two collusion nodes, in which one is sender and the other is receiver, these two nodes will launch the attacks easily. But their given attacking patterns in their simulation study are simple and may be easy to detect, and they do not discuss distributed attacking. P. Kyasanur et al [9] proposes to modify 802.11 MAC protocol to solve the misbehavior of selfish nodes. In their scheme, the receiver will determine the backoff value of sender, so receiver can punish sender by incrementing the backoff value when finding sender has some misbehaviors. G. Noubir [10] describes a type of DoS attacking with power consuming as its target.

They argue that in the RTS/CTS handshaking packets the NAV value is vulnerability for the attacker can utilize this value to estimate the coming transmission event and send a bit data to interfere the normal frame. It will result in the large number of damage and retransmission of normal frame, which consume large energy on the normal side and only with low energy cost on the attacking side. However, this attacking may be easy detected for its obvious conflict with the sending behavior defined in the MAC protocol. Other researches use game theory to deal with the problem of selfish nodes [11]. They design distributed protocol make the nodes converge to Nash equilibrium of bandwidth by allocating a cost to each node before its access to channel.

I. Aad [12] analyzes the DoS attacking to close-loop protocol, such TCP and open-loop protocol, such as UDP. They describe a JellyFish attacking to TCP through a packets disordering, period packets dropping and delay variance jittering to make the TCP functional maladjustment. They also describe a Black Hole attacking to the UDP, in which the nodes in the path will drop packets like a black hole.

For the wired DDoS, papers [13] and [10] describe the shrew attacks and RoQ attacks, which are all concerns to protocol and congestion. Shrew attacking can make TCP go into timeout status and start slow start phase frequently. The RoQ attack is to compromise the protocol vulnerability for the reduction of quality of service.

In this paper, we analyze congestion based DDoS attacking in mobile ad hoc networks for the first time to our knowledge. We figure out the principles of the attacking using network capacity analysis and classify the attacking into four categories: Pulsing attacking, Round Robin attacking, Self-whisper attacking and Flooding attacking. The defense schemes are given including the detection signal of RTS/CTS packets, interference frequency and retransmission times and the ECN marking response mechanism Attack graphs have been used to show multiple attack paths in large scale networks.

They have been proved to be useful utilities for network hardening and penetration testing. However, the basic concept of using graphs to represent attack paths has limitations. In this paper, we propose a new approach, the attack grammar, to model and analyze network attack sequences. Attack grammars are superior in the following areas: First, attack grammars express the interdependency of vulnerabilities better than attack graphs. They are especially suitable for the IDS alerts correlation. Second, the attack grammar can serve as a compact representation of attack graphs and can be converted to the latter easily. Third, the attack grammar is a context-free grammar. Its logical formality makes it better comprehended and more easily analyzed. Finally, the algorithmic complexity of our attack grammar approach is quartic with respect to the number of host clusters, and analyses based on the attack grammar have a run time linear to the length of the grammar, which is quadratic to the number of host clusters.

In this paper, we have proposed a grammar-based approach to modeling and analyzing multi-step network attack sequences. Methods for modelling attacks with PDA are proposed and algorithms for constructing, simplifying and analyzing the attack grammars are demonstrated. The underlying purpose of the new approach is to conquer the limitations of using graphs in previous researches. However, since the algorithm complexity for converting attack grammars to attack graphs is merely linear to the length of the grammar, it is also possible and instructive to construct attack graphs using our method.

One possible future work might be finishing our tool and trying out the attack grammar on realistic networks. Another potential work is to combine our approach to existing IDS applications for alerts correlation. Most cyber-attacks are not single attack actions[18]. They are multi-step attacks composed by a set of attack actions. Although techniques used by attackers can be diverse, attack patterns are generally finite. So we need to find attack steps that are correlated in an attack scenario. By studying the patterns of multi-step cyber attacks, an algorithm is presented for correlating multi-step cyber attacks and constructing attack scenario system based on modeling multi-step cyber attacks. When alerts appear, the algorithm turns them into corresponding attack models based on the knowledge base and correlates them, whether alert or not is based on the weighted cost in the attack path graph and the attack degree of the corresponding host.

And attack scenarios can be constructed by correlating the attack path graphs. Moreover, the model can detect intrusion alerts in real time and revise the attack. A novel attack pattern modeling method was proposed to correlate single attack actions and construct attack scenarios in this paper. By our approach, firstly, the raw alerts are preprocessed. Then, the attack path graphs are constructed by correlated multiple alerts. Based on the attack path graphs, attack scenarios are generated.

The results are forwarded to the security administrator for the intrusion analysis. One of the most important works in the future is to infer the full scope of the attack from the attack scenarios which allows for deeper analysis. Another direction of our future research is to create better and more sophisticated pattern for the individual steps which the attacker takes to compromise a system. In addition we plan to test our approach using data captured from a live network.

A Proactive Test Based Differentiation ract— Low rate DoS attacks are emerging threats to the TCP traffic, and the VoIP traffic in the Internet. They are hard to detect as they intelligently send attack traffic inside the network to evade current router based congestion control mechanisms. We propose a practical attack model in which botnets that can pose a serious threat to the Internet are considered. Under this model, an attacker can scatter bots across the Internet to launch the low rate DoS attack, thus essentially orchestrating the low rate DoS attack that uses random and continuous IP address spoofing, but with valid legitimate IP addresses. It is difficult to detect and mitigate such an attack.

We propose a low rate DoS attack detection algorithm, which relies on the core characteristic of the low rate DoS attack in introducing high rate traffic for short periods, and then uses a proactive test based differentiation technique to filter the attack packets. The proactive test was originally proposed to defend DDoS attacks and low rate DoS attacks which tend to ignore the normal operation of network protocols, but it is tailored here to differentiate the legitimate traffic from the low rate DoS attack traffic instigated by botnets. It leverages on the conformity of legitimate flows, which obey the network protocols. It mainly differentiates legitimate connections by checking their responses to the roactive tests which include puzzles for distinguishing botnets from human users. We finally evaluate and demonstrate the performance of the proposed low rate DoS attack detection and mitigation algorithm on the real Internet traces.

The end host based defense strategy against the low rate TCP DoS attack is randomization of the minimum RTO [2] [23], but it cannot defend against an RoQ attack as it exploits the network element, and not the end host's vulnerability. A few router based defense systems have been proposed to mitigate the low rate DoS attacks. In [24], a modified AQM scheme is proposed to penalize bursty flows, but it lacks accurate identification of the attack flow, and can penalize legitimate bursty short-lived TCP flows. It does not consider the spoofing problem of an RoQ attack. The RoQ attacks can use source and destination IP address spoofing, and they do not have well-defined periodicity, and so schemes like [25]-[27] may not filter these attack packets accurately. The scheme proposed in [28] may not be scalable, and can be deceived by the IP address spoofing, although it can detect any periodic pattern in the flows. The wavelet based approach [5] relies on detecting variability of the incoming traffic rate and outgoing acknowledgments for detecting the low rate TCP DoS attack, but this approach cannot mitigate the RoQ attack which does not try to completely shutdown the competing flows. The extent of the IP spoofing problem in the Internet today is demonstrated in [13], and thus the spoofing issue cannot be ignored. Sarat and Terzis [29] suggested to decrease the buffer size by means of a mathematical analysis based on [30] to expose the attack flow to the AQM schemes like RED-PD, but unfortunately it cannot mitigate the RoQ attack, in which the average rate is lower than the low rate TCP DoS attack. In this authors proposed a detection system to detect new breed of DoS attacks in the Internet known as low rate DoS attacks. Our detection system was shown to detect and mitigate these attacks even if an attacker uses IP address spoofing.

We have evaluated the feasibility of the proposed low rate DoS attack algorithm on real Internet traces. As part of the future work, we are planning to test the effectiveness of the proposed detection system by using the NSF Deterlab test bed.
In this field we can see that lot of work has done and still research work is going on due to new technologies. In Collision attack, the adversary may only need to induce a collision in one octet of a transmission [11]. A minute change in the data portion of the packet will result in a checksum change, hence requiring an expensive exponential back off in some MAC protocols. Error Correcting Codes can be used to tolerate variable levels of corruptions in the messages at any layer. However these error-correcting codes can only work up to a threshold of corruption and they themselves induce additional computational and communicational costs. The attack is based on partial key exposure vulnerability [28] in the RC4 stream cipher. Denial-of-Service attacks [19], and jamming in particular, are a threat to wireless networks because they are at the same time easy to mount and difficult to detect and stop. This paper proposes a distributed intrusion detection system, in which each node monitors the traffic flow on the network and collects relevant, as mobile ad hoc network applications are deployed, security emerges as a central requirement. This paper introduces the wormhole attack [20], a severe attack in ad hoc networks that is particularly challenging to defend against. The wormhole attack is possible even if the attacker has not compromised any hosts.

Network security in a wireless LAN environment is a unique challenge. Whereas wired networks send electrical signals or pulses through cables, wireless signals propagate through the air. Because of this, it is much easier to intercept wireless signals. This extra level of security[] complexity adds to the challenges network administrators. As organizations continue to spend millions of dollars to ensure corporate, customer and financial data is protected, they don't realize wireless devices entering the work place render these millions of dollars of investment potentially obsolete. Hackers [22] armed with tools like hunter killer, Air Jack, the

Evil Twin phishing attack etc. can compromise. What three things about WLANs wireless LANs strike fear into their hearts, and the answers are likely to be similar: Security, Security, and Security [23]. Sure, you want good coverage. Sure, you want to minimize drops. As with any other type of network, an analyzer [24] is one of the first purchases that a wireless network administrator should make. In addition to the traditional functions of protocol analysis and diagnostic tool, wireless network analyzers can be used to address many of the security concerns that may cause to reduce performance of networks. As networking evolves to support both wired and wireless access, securing corporate networks [25] from attack becomes ever more essential. Intel IT is using a new security method to authenticate devices, validate them against security compliance policies, and remediate specific problems before they connect to Intel's networks. There is little wonder why wireless networks have become so ubiquitous the last few years.

Wireless access points [26] are inexpensive, easy to install, and most of all, handy. Although wireless networking can make life a lot easier for your users though, they can also become a security nightmare.Data transfer that relies on carrier sense, which let the nodes detect if other nodes are currently transmitting, are particularly vulnerable to DoS. Link-layer security architecture can prevent this type of attack by detecting unauthorized packets when they are first injected into the network, thus putting a stop to energy and bandwidth waste. TinySec support two different security options: Authenticated Encryption and Authentication only. In authentication encryption mode, data payload is encrypted and the entire packet is secured by a MAC. In contrast, a packet is only secured with Message Authentication Code in authentication mode thereby decreases the power consumption.

### 3. Understanding MAC layer protocols

In this, only CSMA in fading channels is discussed. Predicting the channel condition based on channel state information and regulating the load into the channel accordingly to improve the throughput in faded CSMA channel, In this, finding prediction error of channel state estimation and relevant consequences in load regulated CSMA is not explained and practical settings is also lacking. Carrier Sense Multiple Access with Collision Avoidance (CSMA/CA) [16] is a widespread technique used in modern wireless communication systems to allow multiple users to share the same frequency resource (band). A CSMA/CA compliant system must sense the in-channel power before transmitting. If the measured power is below a certain threshold, meaning that the medium is actually not busy, then the transmission begins. Else, after a random back-off period, another attempt is performed. Common examples of wireless systems using CSMA/CA are Wi-Fi networks (IEEE 802.11 standard) and IEEE 802.15.4 [17] wireless sensor networks (WSN).

We have seen that hybrid MAC protocol, called Z-MAC [31] for Wireless Sensor Network Combined the Strength of Both CSMA and TDMA. Like CSMA, ZMAC achieves high channel utilization under low contention and like TDMA, achieve high channel utilizations under high contention and reduced collision among two-hop neighbor at a low cost.

A distinctive feature of Z-MAC is that its performance is robust to Synchronization error, Slot assignment failure and time varying channel conditions; in the worst case, its performance always falls back to that of CSMA, Z-MAC is implemented in TinyOS. Performance of Z-MAC in Wireless Mess Network and Mobile ad hoc Networks has to be analyzed.

The performance of modern wireless communication systems are strictly related to the adopted medium access mechanism. An effective one, based on the Carrier Sense Multiple Access with Collision Avoidance technique, is here in after investigated. In particular, the performance of a CSMA/CA-based wireless sensor networks, designed for industrial monitoring [13] and performing both cyclic polling and acyclic alarm management, are analyzed. Effects of in-channel interference are also studied. The aim of the paper is to gain, from practical experiments, results and guidelines useful for the measurement and optimization of CSMA/CA-based sensor network performance, when employed for industrial monitoring. In such direction, a suitable IEEE 802.15.4 wireless sensors' tested has been deployed and a high layer protocol for industrial monitoring has been implemented.

In this, one class of applications envisaged for the IEEE 802.15.4 LR-WPAN (low data rate wireless personal area network) standard is wireless sensor networks for monitoring and control applications. In this, an analytical performance model for a network in which the sensors are at the tips of a star topology [14], and the sensors need to transmit their measurements to the hub node so that certain objectives for packet delay and packet discard are met. In this, we first carry out a saturation throughput analysis of the system; i.e., it is assumed that each sensor has an infinite backlog of packets and the throughput of the system is sought. After a careful analysis of the CSMA/CA MAC that is employed in the standard, and after making a certain decoupling approximation, It was identify an embedded Markov renewal process, whose analysis yields a fixed point equation, from whose solution the saturation throughput can be calculated. It was found that with the default back-off parameters the saturation throughput decreases sharply with increasing number of nodes. In this, Manzur Ashraf derive throughput of a threshold based transmission policy, namely load regulated CSMA [15], taking into account the propagation delay of the medium and the offered load at a different probability of the fading channel but analytical framework of CSMA/CA has not yet been analyzed.

In Continuous Channel Access or Exhaustion attack, a malicious node disrupts the MAC protocol, by continuously requesting or transmitting over the channel. This eventually leads a starvation for other nodes in the network w.r.t channel access [32]. One of the countermeasures to such an attack is Rate Limiting as described in [Wood02]. Here, the network ignores excessive requests without sending expensive radio transmissions. This limit however cannot drop below the expected maximum data rate the network has to support. This limit is usually coded into the protocol during the design phase and requires additional logic also. Repeated application of these exhaustion or collision based MAC layer attacks can lead into unfairness. This kind of attack is a partial DoS [32] attack, but results in marginal performance degradation. One major defensive measure against such attacks is the usage of small frames, so that any individual node seizes the channel for a smaller duration only. However the adversary can still cause starvation by frequently requesting for channel, while others go on a random back off. Wireless communications are vulnerable to interception and attack, making it essential that you take every precaution in securing your wireless network. Wi-Fi is an abbreviation for wireless Fidelity [27] and is used to refer generically to any type of wireless network based on the IEEE 802.11 standard. Wi-Fi offers users easy wireless network set-up, access, and use, but because the standard was designed with only limited security capabilities.

Learn about the pervasive nature of wireless communications [32], and discover some security methods that will protect your network against unauthorized external access. Wireless networks have been a ripe target for attacks [30] ever since the emergence and widespread adoption of the 802.11b. Security capabilities of wireless devices can often be confusing and difficult to set up and administer - and have been under constant scrutiny and attack in the trade press.

At national level there is little to report but at international level there are noteworthy contributions, which are taken as base reference for my research work.

## 4. Proposed methodology

We have to construct an system through which we can learn the pattern of attacks and also we can classify then in 3 parts as unknown and unidentified pattern attack, known but un identified patter attack and last is known and identified pattern. It is possible by maintaining database of previous pattern of attacks

We can adopt a system and can generate real time attack and also we can detect there pattern by passing them into an updated Intrusion detection system and those new papttern can be store or accumulate with available previous data base collection of attacks.

The MAC Protocols in current state can be considerably inefficient due to poor channel utilization during dynamic/fluctuating load traffic. The Problem gets worse when the channel is subjected to different kind of vulnerabilities especially malicious attacks which results into drastically increase in data traffic and thereby further decrease the effective channel bandwidth utilization.

In our proposed work we focus our effort to devise an efficient and attack resilient MAC protocol for Computer network.

In this investigation we first survey the various channel allocation schemes and adaptive algorithm dealing with increase data traffic. We have implemented the concept of adaptive contention window size, as per BEB (Binary Exponential Backoff algorithm). Simulation result by Standard Network Simulator, shows that performance of adaptive CSMA/CD is improved comparable to original one. Our survey will also include approaches which sense and adopt to the various threats, this will enabling us to critically investigating the problem of existing techniques to mitigate the challenges in Medium Access Controls. Subsequently We Focus Our Efforts on further improving the channel allocation scheme and include increase in data traffic due to various attacks.

We have to develop an Adaptive, Medium Access Control Protocol using which can learn pattern of attacks and mitigate them.

## 5. Conclusion and Future work.

We know that there are so many possible attacks in wireless and wired networks so it required large database collection for storing patern of all of them and it will also increase complexity of systems as well as we have to put more time to search and match the pattern while attack is performed Identifying difference between sudden increase in traffic and attack is only possible when we have some rules and some previous learned pattern of attacks.

As per our survey work, Assessments of the existing Medium Access Control Protocols with respect to handling increased data traffic caused by malicious Attacks. We aim to develop An Attack Resilient approach for increasing the channel utilization for Medium Access Control Protocol.